\documentclass[prl,reprint]{revtex4-1}
\usepackage{amsmath}    % need for subequations
\usepackage{graphicx}   % need for figures
\usepackage{verbatim}   % useful for program listings
\usepackage{color}      % use if color is used in text
\usepackage{hyperref}   % use for hypertext links, including those to external documents and URLs
\usepackage{esint}

\begin{document}

\title{The liquid helix}
\author{E. Jambon-Puillet$^{1}$, W. Bouwhuis$^{2,3}$, J.H. Snoeijer$^{2}$, D. Bonn$^{1}$}
\affiliation{$^{1}$Institute of Physics, Van der Waals-Zeeman Institute, University of Amsterdam, Science Park 904, 1098 XH Amsterdam, the Netherlands}
\affiliation{$^{2}$Physics of Fluids Group, Faculty of Science and Technology, MESA+ Institute, University of Twente,
7500 AE Enschede, Netherlands}
\affiliation{$^{3}$School of Life Science, Engineering \& Design at Saxion University of Applied Sciences, P.O. Box 70.000, Enschede 7500 KB, The Netherlands}

%\date{\today}

\begin{abstract}
From everyday experience, we all know that a solid edge can deflect a liquid flowing over it significantly, up to the point where the liquid completely sticks to the solid. Although important in pouring, printing and extrusion processes, there is no predictive model of this so-called ``teapot effect''. By grazing vertical cylinders with inclined capillary liquid jets, we here use the teapot effect to attach the jet to the solid and form a new structure: the liquid helix. Using mass and momentum conservation along the liquid stream, we first quantitatively predict the shape of the helix and then provide a parameter-free inertial-capillary adhesion model for the jet deflection and critical velocity for helix formation.
\end{abstract}

\maketitle

When a liquid is poured too slowly from a container, it has the tendency to ``stick'' to the container edge, running down along the container's wall instead of separating from it. To avoid this inconvenience often referred to as the ``teapot effect'', centuries of empirical evidences have taught potters that the design of the container edge, and in particular its sharpness, is of paramount importance. It was however demonstrated only recently by \citet{Duez:2010} that even for rapid inertial flows, the wettability of the surface also plays an unexpectedly important role, and can be used to control liquid flow separation \cite{Dong:2015a,Dong:2015b}. Yet, although the teapot effect has received attention from physicists for decades \cite{Reiner:1956,Keller:1957,Lin:1978,VandenBroeck:1986,VandenBroeck:1989,Kistler:1994,Isshiki:2009,Duez:2010,Dong:2015a,Dong:2015b,Kibar:2017}, a simple quantitative description fully capturing the observations is still lacking. 
From a practical point of view, understanding the teapot effect is of paramount importance not only for designing food containers, but also to better control flows through orifices \cite{Ferrand:2017}, to avoid fouling up the nozzle of inkjet and 3D printers \cite{krichtman:2014} and for polymer extrusion processes where capillary adhesion causes ``sharkskin'' instabilities \cite{Inn:1998}.

\begin{figure*}[bt]
\centering
\includegraphics[width=\textwidth]{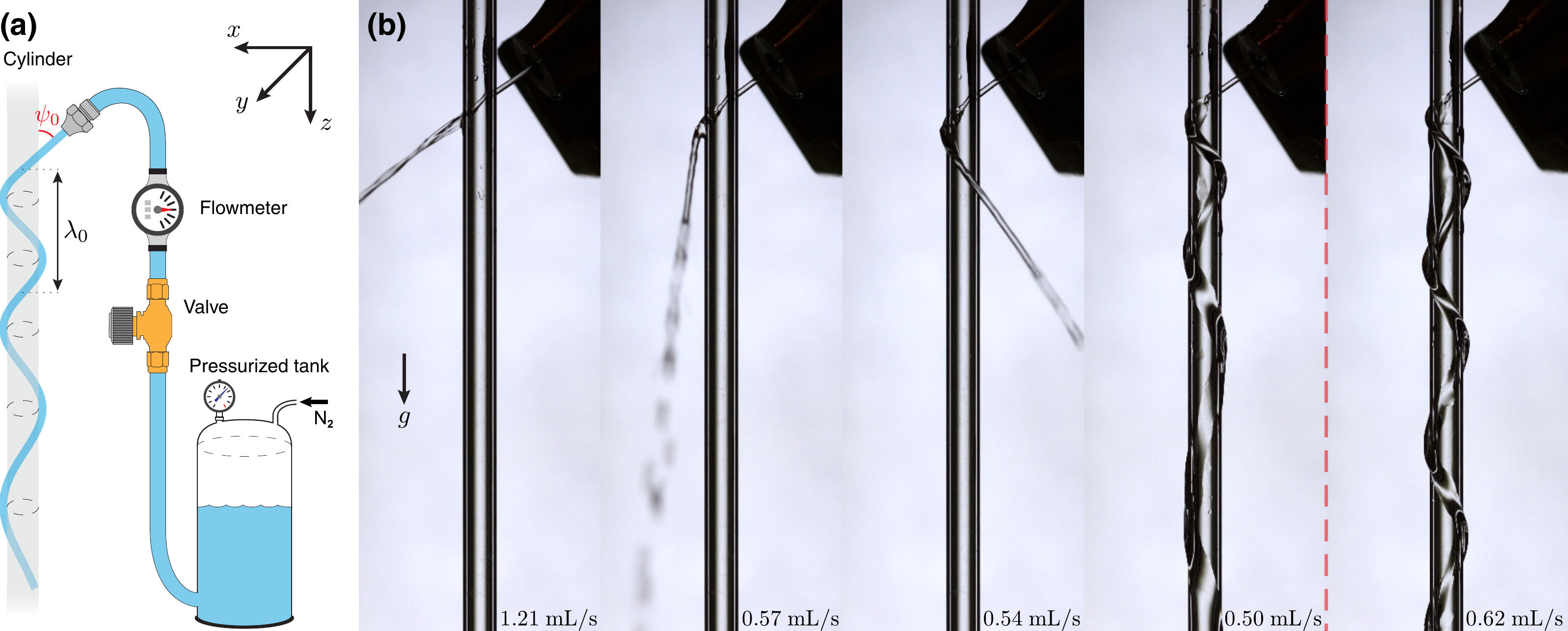}
\caption{\label{fig:Fig1} \textbf{(a)} Schematic of the experiment, indicating the inclination angle $\psi_0$ and the initial helical pitch $\lambda_0$. \textbf{(b)} Side-view pictures of a  sequence of experiments showing the deviation of a $0.5$ mm water jet grazing a $3.0$ mm glass cylinder. The flow rate $Q$ is decreased until the penultimate image and then increased again to illustrate the hysteresis in the sticking transition.}
\end{figure*}

In this Letter, we experimentally investigate the adhesion of capillary water jets to a vertical glass cylinder (Fig. \ref{fig:Fig1}). High speed jets are deflected due to inertial-capillary adhesion, and upon decreasing the flow rate they eventually fold around the cylinder and completely stick to it. The jet then turns into a steady rivulet which flows down the cylinder, forming an elegant novel fluidic structure, the liquid helix, thus transforming an everyday annoyance into an simple way to produce complex patterns analogous to those of ``liquid rope coiling'' that recently received much attention \cite{Ribe:2012}. We first investigate the rivulet helical trajectory over a wide range of geometrical parameters. We then look into the high velocity regime when the jet is bent by the cylinder but still separates and identify the critical velocity to form a liquid helix. All these results can be accounted for using momentum conservation on the liquid stream, both for the helix shape and the jet deflection. In particular, the proposed model is the first to actually predict a sticking transition, and its scaling laws are in excellent agreement with experiments.

%
%
%Previous studies of the teapot effect are mostly theoretical in their approach, providing either partial models \cite{Keller:1957,Lin:1978,Vanden‐Broeck:1986,Vanden‐Broeck:1989,Isshiki:2009}, computational fluid dynamics simulations \cite{Kistler:1994,Kibar:2017} or scaling arguments \cite{Duez:2010,Dong:2015a}. A recent inertial-capillary model \cite{Duez:2010} based on scaling arguments for the momentum balance qualitatively captures all their experimental trends, but is not quantitative since the scaling arguments do not allow to estimate the prefactors. In addition what is perhaps the central question about the teapot effect, namely at which speed the liquid jet sticks or unsticks from the solid, remains unanswered so far. Here the experiments of Duez et al \cite{Duez:2010} disagree with more recent experiments by \citet{Dong:2015a}; both find different dependences of the critical speed for the fluid to stick on the curvature of the solid edge. %

Our helix experiment is shown schematically in Fig. \ref{fig:Fig1}(a). A jet inclined by an angle $\psi_0$ with respect to the vertical is generated by flowing water (density $\rho=1$ g/cm$^3$, viscosity $\eta=1$ cP, surface tension $\gamma=72$ mN/m) from a pressurized tank through a nozzle (bore diameter $0.2<D_j\:(\mathrm{mm})<1.5$). The volumetric flow rate $Q$ is kept constant (controlled by a precision valve and measured using a flow meter). The jet is impacted on a vertical cylinder made of glass (contact angles $\theta\approx 30^\circ$ with fluctuations between cylinders) or teflon ($\theta\approx 90^\circ$) with a diameter $1.05<D_c\:(\mathrm{mm})<14.4$ (See Supplementary Material \footnote{See Supplementary Material at [URL will be inserted by publisher] for details on the experimental setup, derivation of the models and in depth analysis of the inertial-capillary adhesion model. It includes Refs \cite{birdi:1989,Carre:1995,Lamberti:2012,yasuda:1994,Bouwhuis:2015,Bush:2004,Celestini:2010}.}). As the degree of overlap between the jet and the cylinder is a critical parameter \cite{Kibar:2017}, we use a linear stage to translate the nozzle until it barely touches the cylinder. Fig. \ref{fig:Fig1}(b) shows photographs of an experiment, in which the flow rate is decreased and increased again. The pictures show that as the flow rate $Q$ is decreased, the water jet is increasingly bent by the glass cylinder until at a critical flow rate it completely sticks to the cylinder, forming a helical rivulet. This sticking transition is hysteretic: increasing the flow rate again does not cause the immediate breakdown of the helix. For all our experiments, the jet Reynolds and Froude numbers are quite high: $360< \mathrm{Re}=\rho U_0 D_j / \eta <6600$ and $7< \mathrm{Fr}= U_0 / \sqrt{g D_j} <308$ with $U_0=4Q/\pi D_j^2$ the initial jet speed. The initial phase of the jet-sticking will thus be governed by inertia, though it will turn out that viscosity and gravity affect the helix after a couple of revolutions.

\begin{figure}[bt]
\centering
\includegraphics[width=\columnwidth]{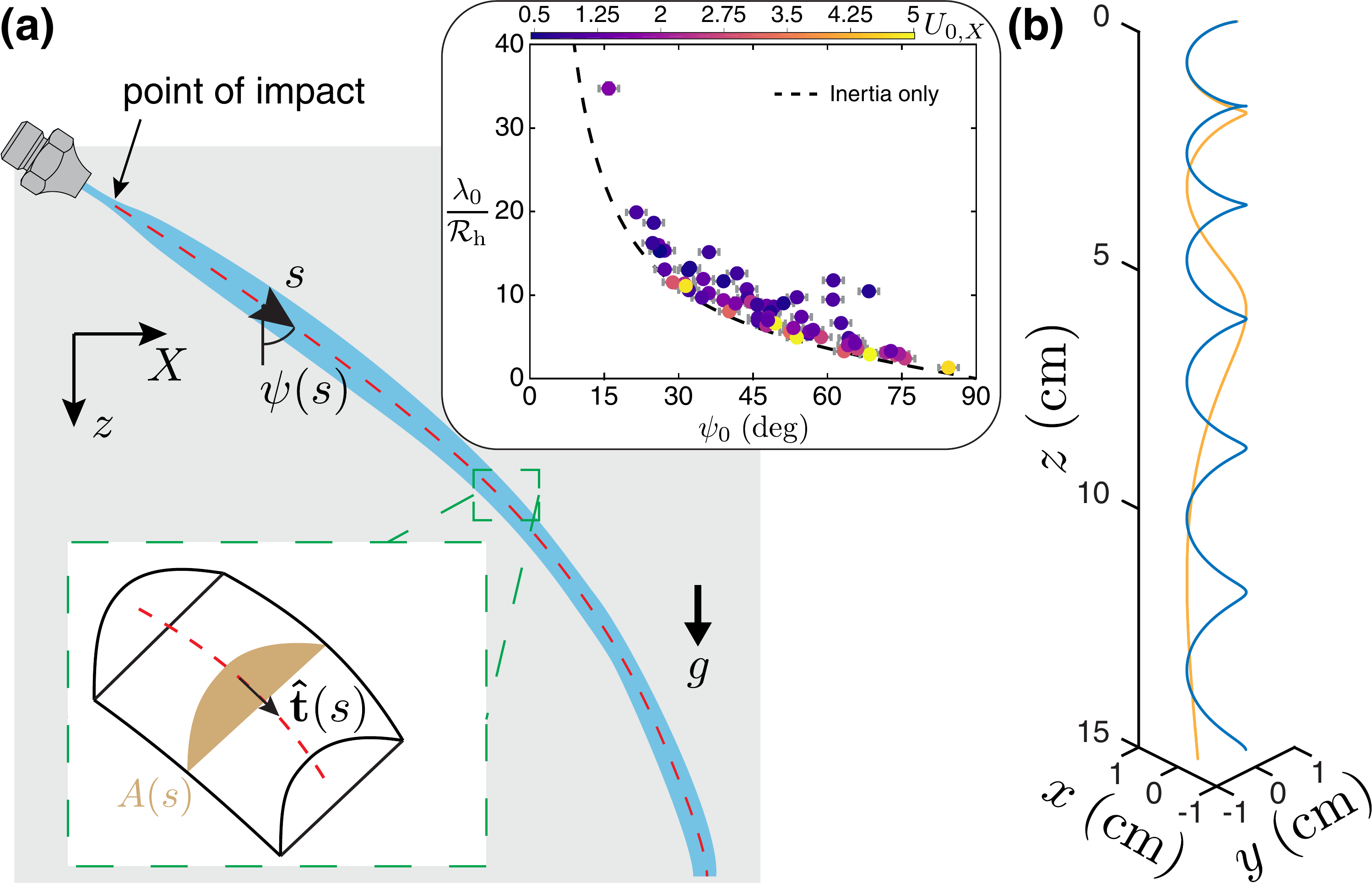}
\caption{\label{fig:Fig2} \textbf{(a)} Sketch of the (unwrapped) rivulet model. 
%The jet impacts a (flat) solid surface and turns into a rivulet whose centerline (drawn as a red dashed line) is parametrized through its arc length $s$ and its angle with respect to the vertical axis $\psi(s)$. 
The impacting jet turns into a rivulet whose centerline (drawn as a red dashed line) is parametrised by the arclength $s$ and the angle with respect to the vertical axis $\psi(s)$. 
\textit{Inset:} Initial helix pitch $\lambda_0/\mathcal{R}_{\mathrm{h}}$ as a function of the jet inclination angle $\psi_0$, for varying $D_j$, $D_c$ and initial velocity $U_{0,X}$ (color-coded in m/s). Dashed line is the inertial prediction $\lambda_0=2\pi\mathcal{R}_{\mathrm{h}}/\tan\psi_0$. \textbf{(b)} Theoretical helix shape including the effect of gravity with (orange curve) and without (blue curve) viscous effects for typical experimental parameters. 
%Experimental parameters $D_c=10$ mm, $D_j=0.8$ mm, $\psi=64^\circ$ and $U_0=2.1$ m/s. Fitting parameter $C=11$.
}
\end{figure}

\paragraph{The shape of the helix.--}~We first focus on the helical rivulet regime, for which the jet sticks completely to the cylinder. Neglecting the small thickness variations, we assume that the fluid stream describes a helical motion with a constant helix radius $\mathcal{R}_{\mathrm{h}}=\left( D_c+D_j \right) / 2$. The rivulet trajectory can then be parametrised by the rivulet arc length $s$ and its local angle with respect to the vertical $\psi(s)$. It is insightful to ``unwrap" the trajectory on an effective Cartesian plane ($X$,$z$) shown on Fig. \ref{fig:Fig2}(a). The azimuth $\phi$ is then replaced by $X=\mathcal{R}_{\mathrm{h}}\phi$ and the tangent vector becomes planar (see SM \cite{Note1}). The problem then becomes mathematically equivalent to finding the trajectory of the rivulet formed by the impact of a jet of vanishing incidence on a flat plate. Given the large $\mathrm{Re}$ and $\mathrm{Fr}$, we anticipate the initial revolution to be dominated by inertia. The initial $z$-momentum is unchanged, while the $x$-momentum is transferred to the orthoradial direction $X$ (once unwrapped). Without viscous friction or gravity, the unwrapped rivulet trajectory is trivially a straight line; this corresponds to an helix of constant pitch $\lambda=2\pi\mathcal{R}_{\mathrm{h}}/\tan\psi_0$ once wrapped around the cylinder. In the inset of Fig. \ref{fig:Fig2}(a), we compare this prediction to the experimentally observed initial pitch $\lambda_0$ [defined in Fig. \ref{fig:Fig1}(a)]. Indeed, the inertial prediction accurately describes the initial pitch $\lambda_0$, except for the slowest jets.

%Since we expect an inertia dominated problem, we only consider it as a first approximation. As such, the inertial-capillary adhesion transfers all of the initial jet momentum to the helical rivulet; the $z$ momentum is unchanged while the $x$ momentum is transferred to the orthoradial direction X (once unwrapped). Because there is neither friction nor gravity, the rivulet speed remains constant (equal to the initial jet speed) and the unwrapped rivulet trajectory is trivially a straight line which corresponds to an helix of constant pitch $\lambda=2\pi\mathcal{R}_{\mathrm{h}}/\tan\psi_0$ once wrapped around the cylinder. We plot in inset of Fig. \ref{fig:Fig2}(a) the initial pitch $\lambda_0$ of all our helices (see Fig. \ref{fig:Fig1}(a)) and compare it to this prediction. Despite its simplicity, this zero-order model captures the beginning of the experimental trajectory quite well, except for the slowest jets. However, the actual pitch is clearly not constant and increases the further the helix goes down (Figs. \ref{fig:Fig1}(b) and \ref{fig:Fig3}).  After a few turns the rivulet has lost most of its orthoradial momentum and the liquid only flows downward, unless it starts a meandering like motion \cite{rivulet1,rivulet2} probably triggered by surface defects.

\begin{figure}[t]
\centering
\includegraphics[width=\columnwidth]{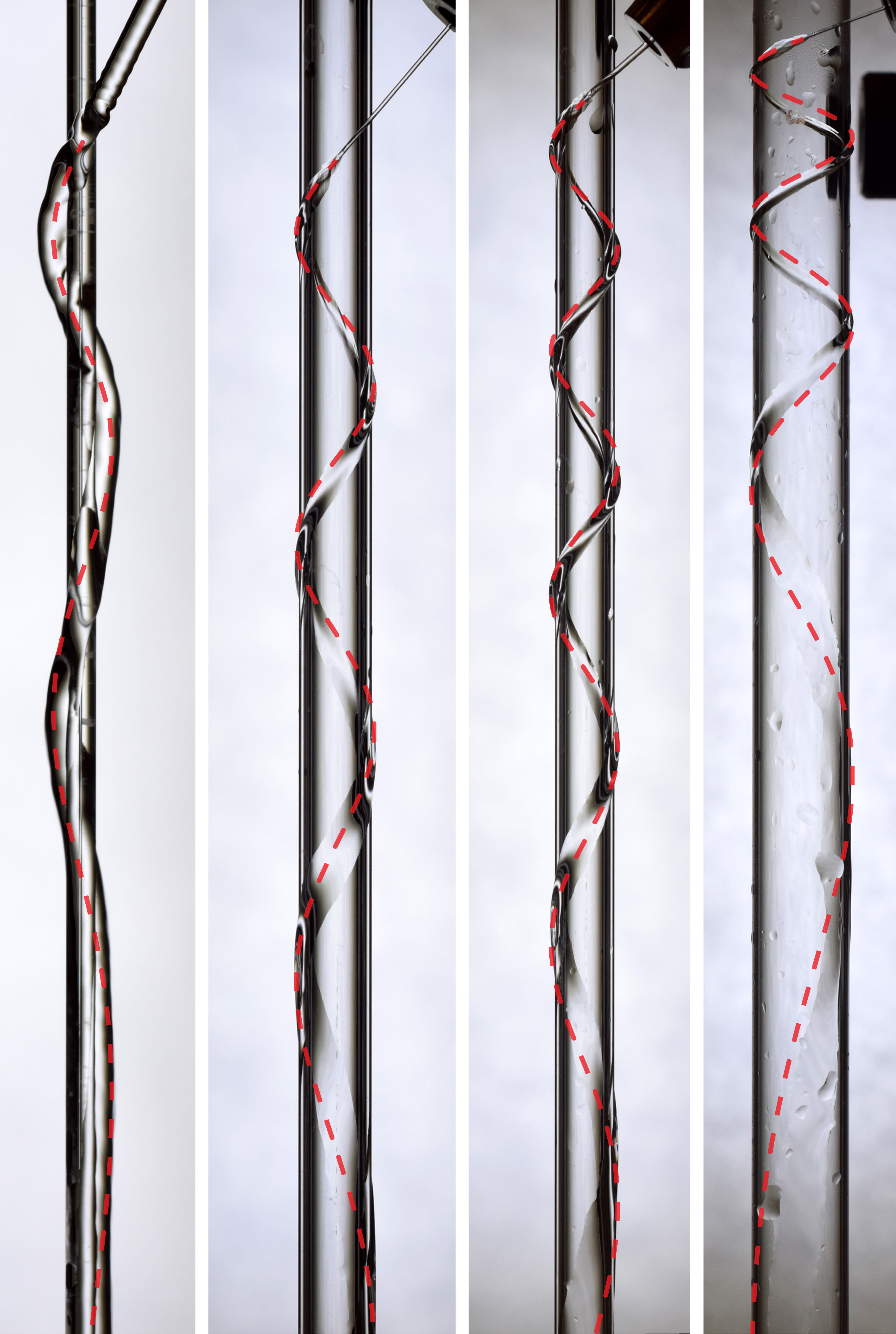}
\caption{\label{fig:Fig3} Comparison between experiments and theory for a range of experimental parameters. From left to right: $D_c$ = 1, 5, 5, 10 mm, $D_j$ = 1, 0.3, 0.5, 0.3 mm; $\psi_0$ = 26.3, 40.3, 47.8, 68.6 deg; $U_0$ = 1.0, 4.9, 3.0, 5.4 m/s; $C$ = 15, 11.5, 7, 9.5.
}
\end{figure}

However, the actual pitch is clearly not constant and increases as the helix goes down (Figs. \ref{fig:Fig1} and \ref{fig:Fig3}). After a few turns the rivulet has lost most of its orthoradial momentum and the liquid only flows downward. Introducing gravity into the inertial description indeed stretches the helix, but only by a negligible amount (see Fig. \ref{fig:Fig2}(b), blue curve). Instead, a quantitative description of the helix calls for both gravity \emph{and} viscous friction. In the spirit to the analysis of hydraulic jumps \cite{Watson:1964,Wang:2013} and meandering rivulets \cite{rivulet1,rivulet2}, we therefore perform a momentum balance on an infinitesimal portion of the rivulet [see Fig. \ref{fig:Fig2}(a)] including gravity, viscous friction and the inertial-capillary adhesion force. At steady-state, the flux $Q=AU$ is constant along the helix, where we introduced $U(s)$ as the mean rivulet velocity averaged over the cross-sectional area $A(s)$. If we further introduce unit vectors along the rivulet, $\boldsymbol{\hat{\mathbf{t}}}(s)$, and normal to the cylinder $\boldsymbol{\hat{\mathbf{n}}}(s)$, the steady momentum reads:

\begin{equation}
\rho Q \frac{\mathrm{d}(U \boldsymbol{\hat{\mathbf{t}}})}{\mathrm{d}s}=W \left( \tau \, \boldsymbol{\hat{\mathbf{t}}} - \Delta P \, \boldsymbol{\hat{\mathbf{n}}} \right)+\rho A\mathbf{g}.
\label{eq:momgen}
\end{equation}
%\begin{equation}
%\rho Q \frac{\mathrm{d}\vec{U}(s)}{\mathrm{d}s}=\vec{\tau}(s) W(s)+\rho A(s)\vec{g}+\oint_{\mathcal{C}} \Delta P \boldsymbol{\hat{\mathbf{n}}}\mathrm{d} u.
%\label{eq:momgen}
%\end{equation}
In this expression $\tau$ is the wall shear stress, while $\Delta  P$ is the difference of pressure between the upper and lower side of the jet, both averaged over the rivulet width $W(s)$ (see SM \cite{Note1}). 

So far, Eq.~\eqref{eq:momgen} is without approximations. The inertial-capillary adhesion force is encoded in the pressure difference $\Delta P$. In the regime where a helix forms, however, $\Delta P$ will be balanced by the centrifugal acceleration along $\boldsymbol{\hat{\mathbf{n}}}$, but this does not affect the shape of the helix. To estimate the wall shear stress, we assume a two dimensional parabolic flow such that $\tau\approx -3 \eta U/h$, with $h(s)$ the rivulet thickness at the centerline. This is complemented by the geometric assumptions $A\approx Wh$ and $W(s)\approx D_j$ (constant width) such that $\tau W=-3C\eta D_j^2 U^2/Q$ with $C$ a form factor encompassing the three aforementioned assumptions that we consider constant along the stream.
%Projecting Eq. \eqref{eq:momgen} in the full helical geometry shows that the normal adhesion term is balanced by the centrifugal acceleration and that the helix shape solely comes from the momentum balance in the (unwrapped) $X$ and $z$ direction \cite{Note1}. Including mass conservation $Q=U(s)A(s)$, we thus only have three equations for five unknowns ($U(s)$, $\psi(s)$, $A(s)$, $W(s)$, $h(s)$). The problem is thus underdetermined and we must make assumptions on the rivulet shape to close the system of equations. We therefore assume that the cross section can be written as $A(s)=C W(s) h(s)$ with $C$ a form factor constant with respect to $s$ and that the rivulet width is constant and equal to the initial jet diameter $W(s)= D_j$. 
With this, the momentum balance [Eq. \eqref{eq:momgen}] takes the form (cf. SM \cite{Note1}):
\begin{align}
&\frac{\mathrm{d}U}{\mathrm{d}s} = -\frac{48\eta C}{\pi^2\rho D_j^2}\left(\frac{U}{U_0}\right)^2+\frac{g \cos\psi}{U},\label{eq:hel1}  \\
&\frac{\mathrm{d}\psi}{\mathrm{d}s} = - \frac{g\sin\psi}{U^2}, \label{eq:hel2}
\end{align}
once projected along the rivulet in the unwrapped $(X,z)$-plane. The helix shape is then extracted from $\psi(s)$ as $\mathrm{d}z/\mathrm{d}s=\cos\psi(s)$ and $\mathrm{d}X/\mathrm{d}s=\sin\psi(s)$. We numerically integrate Eqs. \eqref{eq:hel1},\eqref{eq:hel2}, with initial conditions ($U_0$, $\psi_0$) and wrap the trajectory around the cylinder to obtain the helix shape. 

Figure~\ref{fig:Fig2}(b) compares calculated helix shape with and without viscous effects using typical experimental parameters. It clearly shows that both gravity and viscosity are necessary to quantitatively account for the experiments: the pitch increases significantly over a few turns and the rotation slows down and eventually stops. The direct agreement with experiment is excellent (Fig. \ref{fig:Fig3}), with $C\sim 10$ as the only adjustable parameter that does not vary much for most of our experimental conditions. The helix shape could therefore be tuned by controlling the friction through the fluid viscosity. However, rotating the cylinder breaks the helix and coats the cylinder with a thin film.

\begin{figure}[bt]
\centering
\includegraphics[width=\columnwidth]{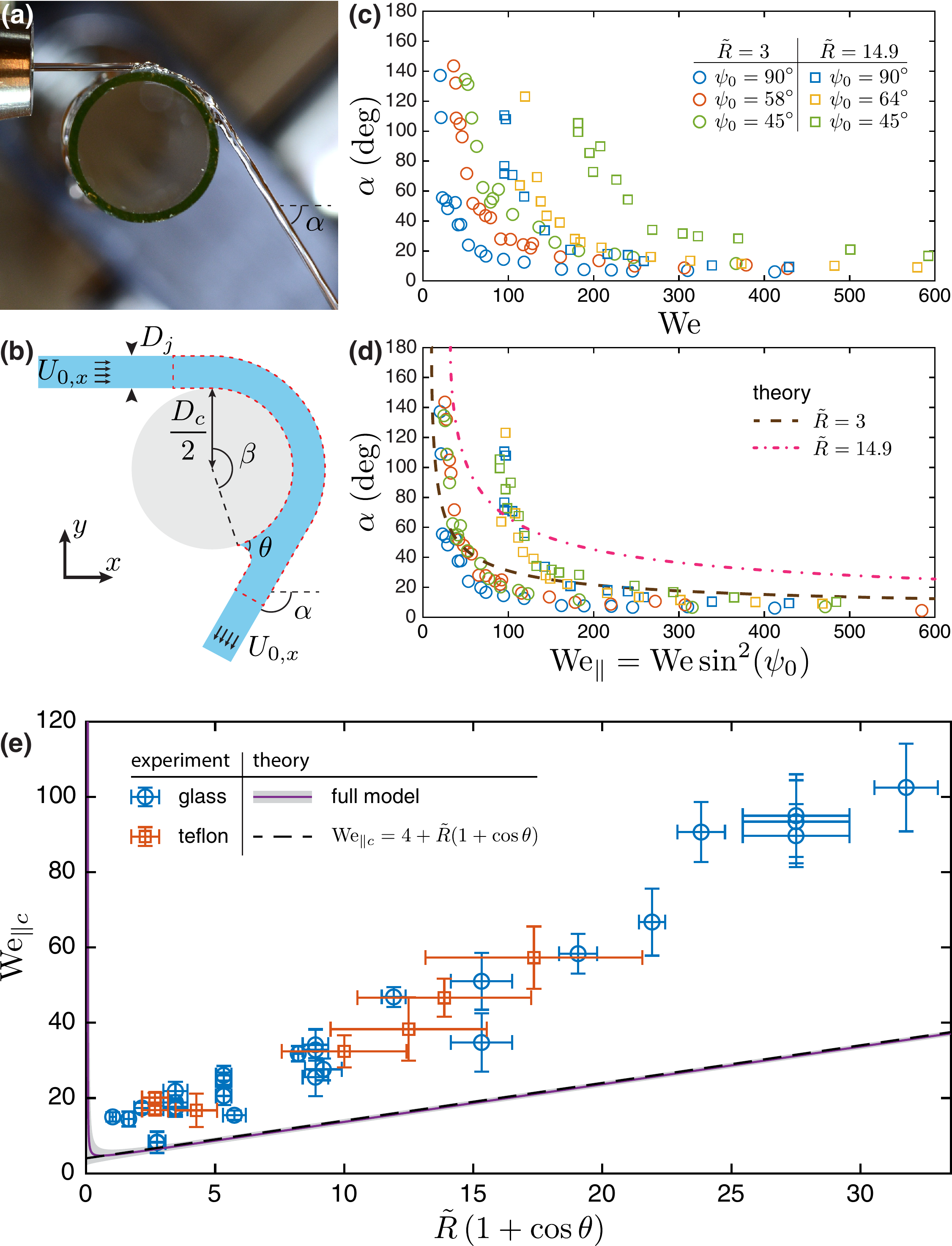}
\caption{\label{fig:Fig4} \textbf{(a)} Top view of a typical experiment ($D_j=0.5$ mm, glass cylinder $D_c=10$ mm, $Q=0.77$ mL/s, $\psi_0=91.5^\circ$). \textbf{(b)} Schetch of the 2D model defining the geometric parameters. The control volume for which we consider the momentum balance is indicated by the red dashed line. \textbf{(c)} Jet deviation angle $\alpha$ as a function of the Weber number $\mathrm{We}$ for different inclination angles $\psi_0$ and dimensionless cylinder radii $\tilde{R}=D_c/(2D_j)$ (glass, $\theta=32\pm 15^\circ$). \textbf{(d)} Same data plotted as a function of the parallel Weber number $\mathrm{We}_{\parallel}=\mathrm{We}\sin^2(\psi_0)$. Dashed lines are results from our (parameter-free) theory [Eqs. (S9)(S10)]. \textbf{(e)} Critical Weber number $\mathrm{We}_{\parallel c}$ as a function of $\tilde{R}\left(1+\cos\theta\right)$. All experimental parameters ($D_j$, $D_c$, $\theta$ and $\psi_0$) are varied. The purple solid curve is the full numerical solution [Eqs. (S9)(S10)] (the grey area represents the variations as $\theta$ is varied from $0^\circ$ to $179^\circ$). The black dashed line is the asymptotic expansion in the large $\tilde{R}$ limit.}
\end{figure}

\paragraph{Critical speed for helix formation.--}~Now that we understand the shape of the helical rivulet, we aim to describe how the jet sticks to the cylinder. In the experiment we measure the jet deviation angle $\alpha$ with respect to the incident jet as we decrease the flow rate $Q$ from top-view pictures [Fig. \ref{fig:Fig4}(a)], and vary the jet size $D_j$, cylinder size $D_c$, inclination angle $\psi_0$ and contact angle $\theta$. Since the jet velocity is higher here than in the helix regime, we fully neglect gravitational and viscous effects. The relevant dimensionless numbers for the experiment are therefore the Weber number $\mathrm{We}=\rho U_0^2 D_j/\gamma$, the dimensionless cylinder radius $\tilde{R}=D_c/(2D_j)$, the contact angle $\theta$ and the inclination angle $\psi_0$.

Fig. \ref{fig:Fig4}(c) shows $\alpha$ as a function of $\mathrm{We}$ for two representative dimensionless cylinder radii $\tilde{R}$ and various inclination angles $\psi_0$ (for these glass cylinders, $\theta=32\pm 15^\circ$). In all cases, the jet deviation is very small at high speeds ($\alpha\sim 5^\circ$) and gradually increases up to a complete overturn ($\alpha=180^\circ$) as the speed is decreased. Once the overturn is reached, the jet sticks to the cylinder and forms the helical rivulet. We observe that larger cylinders, smaller inclination angles and lower contact angles result in stronger jet deviations. In fact, the dependence on $\psi_0$ can be scaled out, by plotting the same data as a function of the Weber projected in the orthoradial direction, i.e. $\mathrm{We}_{\parallel}=\mathrm{We}\, \sin^2 \psi_0$. The collapse shown in Fig. \ref{fig:Fig4}(d) suggests that the problem is effectively two-dimensional, and can be understood from a projection in the horizontal plane [Fig. \ref{fig:Fig4}(b)]. Our experimental findings on the helix qualitatively agree with earlier experiments on fluids flowing from a solid disk \cite{Duez:2010}. Quantitatively, however, our results are different: our experiments show that the jet can make a complete U-turn, and it is only at this point that the sticking transition happens, whereas the previous experiment put the maximum deflection at $\approx 85^\circ$. Another important difference is observed for the sticking transition. In Fig. \ref{fig:Fig4}(e) we plot the critical speed $\mathrm{We}_{\parallel c}$ for all our experiments, and reveal a linear dependence with $\tilde{R}$. This is contrasted with the scaling $\sim \tilde{R}^2$, initially suggested in \cite{Duez:2010}.

%\begin{figure}[bt]
%\centering
%\includegraphics[width=\columnwidth]{Fig3.pdf}
%\caption{\label{fig:Fig3} Critical Weber number $\mathrm{We}_{\parallel c}$ as a function of the dimensionless tube radius $\tilde{R}$ for all our experiments ($D_j$, $D_t$ and $\psi$ are varied). The solid line is the full numerical solution of equations \eqref{eq:xcomp}\eqref{eq:ycomp} with $\theta=30^\circ$, the dashed line is the analytical asymptotic expansion. \red{I can maybe add the uncertainty on $\theta$ for the theoretical curves}}
%\end{figure}

To rationalize these experimental results we now develop an inertial-capillary adhesion model, for the case where the jet separates from the cylinder [Fig. \ref{fig:Fig4}(a,b)]. We return to the momentum conservation Eq. \eqref{eq:momgen}, and make use of the fact that in this regime one can neglect gravity and viscosity. By consequence, both $U=U_0$ and $\psi=\psi_0$ will remain constant, as can be inferred from Eqs. \eqref{eq:hel1},\eqref{eq:hel2} and therefore $A=A_0$ (mass conservation). Hence, we can integrate Eq. \eqref{eq:momgen} along the arc-length as $\rho A_0 U_0^2 \left( \boldsymbol{\hat{\mathbf{t}}}_{\rm out} - \boldsymbol{\hat{\mathbf{t}}}_{\rm in} \right)= - \int \mathrm{d}s\, \Delta P\, W\,\boldsymbol{\hat{\mathbf{n}}}$. This gives the momentum balance for the control volume indicated by the dashed line Fig. \ref{fig:Fig4}(b). Since $\boldsymbol{\hat{\mathbf{n}}}$ is normal to the cylinder, it is natural to project both $\boldsymbol{\hat{\mathbf{t}}}$ and $\mathrm{d}s=\mathrm{d}s_\parallel/\sin\psi_0$ onto the horizontal ($x$,$y$)-plane. As shown in the SM~\cite{Note1} this renders the problem two-dimensional, based on an effective velocity $U_{0,x}=U_0 \sin \psi_0$. This explains the collapse of the deviation angle $\alpha$ as a function $\mathrm{We}_{\parallel}$ given in Fig. \ref{fig:Fig4}(d), and the similarity with the experiment of \cite{Duez:2010}. In the remainder, we therefore continue with a two-dimensional model and assume $A= W^2 = D_j^2$.

%
%Under the simplifying geometric assumptions of a square constant cross section $W(s)=h(s)=D_j$ and a circular meniscus at the separation point (defined by the angle $\beta\neq \alpha$), Eq. \eqref{eq:momgen} (without gravity and friction) can be integrated along the arc-length and projected in the horizontal ($x$,$y$) plane yielding a 2D control volume analysis equivalent to the situation shown in Fig. \ref{fig:Fig4}(b) with $U_{0,x}=U_0 \sin\psi_0$ as initial velocity \cite{Note1}; thus explaining the good collapse of the 

To obtain a quantitative prediction for the jet deflection and the critical speed $\mathrm{We}_{\parallel c}$, we need to evaluate pressure $\Delta P$ on the upper and lower side of the liquid stream. The free surfaces are subjected to the Laplace pressure which can be integrated analytically along the rivulet. By contrast, the pressure on the solid boundary is of hydrodynamic (inertial) origin \cite{Duez:2010}: the bending of the streamlines creates a depression inside the liquid, and gives rise to an adhesive force \cite{Keller:1957,VandenBroeck:1986,VandenBroeck:1989} (sometimes called Coand\v{a} effect). One can compute this dynamic pressure as $-\rho\int_{D_c/2}^{D_c/2+D_j} \frac{u(r)^2}{r} \mathrm{d}r$, based on the velocity $u(r)$ inside the jet (with $r$ the radial coordinate). For our large cylinders ($\tilde{R} \gg 1$), we can consider concentric circular streamlines with $u(r)\sim 1/r$ \cite{Lhuissier:2012}. This profile differs notably from the inviscid flow around a sharp bend ($\tilde{R} \ll 1$) which approaches $u(r)\sim r^{-1/2}$ \cite{VandenBroeck:1986,VandenBroeck:1989}, and hence our analysis is expected to be valid only for $\tilde{R} \gtrsim 1$.
%Projecting the momentum balance in the $x$ and $y$ direction yields Eqs. (S9)(S10)\cite{Note1}). 

%\begin{widetext}
%\begin{align}
%\left[\mathrm{We}_{\parallel}-\mathrm{We}_{\parallel}\mathcal{G}(\tilde{R})-\frac{1}{\left(1+\tilde{R}\right)}\right] + \left[2-\mathrm{We}_{\parallel}\right]\cos\alpha &+ \left[\mathrm{We}_{\parallel}\mathcal{G}(\tilde{R})-\frac{\tilde{R}}{\left(1+\tilde{R}\right)}\right]\cos\beta 
%+ \cos(\beta+\theta) = 0, \label{eq:xcomp} \\
%\left[2-\mathrm{We}_{\parallel}\right]\sin\alpha &+ \left[\mathrm{We}_{\parallel}\mathcal{G}(\tilde{R})-\frac{\tilde{R}}{\left(1+\tilde{R}\right)}\right]\sin\beta + \sin(\beta+\theta) = 0. \label{eq:ycomp}
%\end{align}
%\end{widetext}
%with $\mathcal{G}(\tilde{R})=\left(\frac{1}{2}+\tilde{R}\right) \left(\tilde{R}\left(1+\tilde{R}\right)^2\left[\ln(1+\frac{1}{\tilde{R}})\right]^2\right)^{-1}$ a dimensionless geometrical factor that encodes the velocity profile dependence.

The above formulation allows a parameter-free calculation of the sticking transition (see SM \cite{Note1}) resulting in Eqs. (S9),(S10). Importantly, by numerically solving for the jet deviation $\alpha$, we for the first time provide a theory that captures the emergence of a minimal speed $\mathrm{We}_{\parallel c}$ for flow separation: the momentum balance admits two branches of solutions, both observed in the experiments, that annihilate through a saddle node bifurcation at $\mathrm{We}_{\parallel c}$ and $\alpha=180^\circ$, in close agreement with experiments (see SM \cite{Note1}). The prediction for $\alpha$ is plotted in Fig. \ref{fig:Fig4}(d) without any adjustable parameters. For small cylinders ($\tilde{R}\lesssim 5$), the calculated deviation angles quantitatively matches the experimental data, while for larger $\tilde{R}$ the agreement is only qualitative. Finally, the model resolves how the critical speed depends on \emph{all} the parameters of the problem. The value of $\mathrm{We}_{\parallel c}$ can be computed analytically through an asymptotic expansion around the critical point (see SM \cite{Note1}) and for $\tilde R \gg 1$ it reveals that $\mathrm{We}_{\parallel c}\approx 4+\tilde{R}\left(1+\cos\theta\right)$. We plot in Fig. \ref{fig:Fig4}(e) $\mathrm{We}_{\parallel c}$ as a function of $\tilde{R}\left(1+\cos\theta\right)$ for all our data (different $D_j$, $D_c$, $\psi_0$ and $\theta$), the full numerical solutions of the model for all possible $\tilde{R}$ and $\theta$ (solid curve with grey area) and the asymptotic analytical solution (dashed line). Both the data and the full model collapse, indicating that this simple scaling is able to capture the physics of the ``teapot effect''.
Our result captures both the wettability dependence $\left(1+\cos\theta\right)$ already observed \cite{Duez:2010,Dong:2015a}, as well as the linear dependence on the solid curvature $\tilde{R}$; it therefore settles the discussion of whether the dependence of the critical speed on the radius of the jet should be quadratic (\citet{Duez:2010}) or linear (\citet{Dong:2015a}). Quantitatively, the slope of the linear dependence is roughly a factor two off [Fig. \ref{fig:Fig4}(e)], which we attribute to the simplifying geometric assumptions of the jet's cross-section. 
Calculating the full geometry of the jet goes well beyond the scope of the present contribution as we expect it to be possible only through computational fluid dynamic simulations \cite{Kibar:2017}.  

In summary, we have studied the sticking of inertial-capillary flows to solids, also known as the ``teapot effect'' by grazing vertical cylinders with liquid jets. We have shown that unlike in the pouring configuration, once the jet completely sticks to the solid in our setup, it forms a liquid helix whose intricate shape depends on the jet initial speed and geometry. We then looked at the adhesion itself and how it impacts the jet when it still separates from the cylinder. Using a detailed momentum balance on the rivulet/jet we are able to accurately recover the observed trajectory of our liquid helices using a single fitting parameter. Moreover, we improved the inertial-capillary adhesion scaling analysis, and derived a parameter-free model that, for the first time, predicts the sticking transition and captures experimental observations semi-quantitatively.

\begin{acknowledgments}
We are grateful to C. Coulais for suggesting the equivalence with the unwrapped problem. E.J-P. thanks Shell Global Research for funding
\end{acknowledgments}

%\bibliography{bibspijet}
%merlin.mbs apsrev4-1.bst 2010-07-25 4.21a (PWD, AO, DPC) hacked
%Control: key (0)
%Control: author (8) initials jnrlst
%Control: editor formatted (1) identically to author
%Control: production of article title (-1) disabled
%Control: page (0) single
%Control: year (1) truncated
%Control: production of eprint (0) enabled
%

\end{document}

\title{Supplementary Materials for ``The liquid helix''}
\author{Etienne Jambon-Puillet$^{1}$, Wilco Bouwhuis$^{2,3}$, Jacco H. Snoeijer$^{2}$, Daniel Bonn$^{1}$}
\affiliation{$^{1}$Institute of Physics, Van der Waals-Zeeman Institute, University of Amsterdam, Science Park 904, Amsterdam, the Netherlands}
\affiliation{$^{2}$Physics of Fluids Group, Faculty of Science and Technology, Mesa+ Institute, University of Twente,
7500 AE Enschede, Netherlands}
\affiliation{$^{3}$School of Life Science, Engineering \& Design at Saxion University of Applied Sciences, P.O. Box 70.000, Enschede 7500 KB, The Netherlands}

\begin{abstract}
In this supplementary document, we provide technical details on the experiment and the modelling presented in the main text. The latter involves the detailed derivation and analysis of the equations for the helix shapes, and tests of some of the assumptions.
\end{abstract}

\date{\today}
\maketitle

\section{Experimental details}
We use Schlick nozzles (model 629) with bore diameters $D_j=0.3,0.4,0.5,0.8,1.0,1.5$ mm and tap water (whose interfacial tension was checked with pendant-drop tensiometry, Kruss EasyDrop) to produce our jets. The flow rates are measured with a Bio-Tech mini turbine flowmeter that we calibrated. We rinsed the cylinders with ethanol or acetone and then water before each experiment. All our glass cylinders are straight with a constant diameter ($\pm 1\%$). For teflon surfaces we used a full cylinder ($D_c=4.20$ mm) and wrapped teflon tape around cylinders of glass or steel. We measured the advancing $\theta_a$ and receding $\theta_r$ contact angle of cylinder from optical pictures of the meniscus after having pushed or pulled the cylinder from a water bath. For the teflon tape, we measured the same angles on sessile drops after injection or withdrawal of liquid (Kruss EasyDrop). We show in Table \ref{tab:tube} the average contact angle $\left(\theta_a+\theta_r \right) /2$ and the half hysteresis $\left(\theta_a-\theta_r \right) /2$ for our cylinders. The value and significant hysteresis observed are in agreement with previous studies using these substrates \cite{birdi:1989,Carre:1995,Lamberti:2012,yasuda:1994}. This uncertainty on the contact angle results in the horizontal error bars on Fig. 4(e).

\begin{table*}[t]
\centering
\begin{ruledtabular}
\begin{tabular}{ccccccccccc} 
Material & \multicolumn{7}{c}{glass} & \multicolumn{3}{c}{teflon} \\
\cline{2-8}\cline{9-11}
$D_c$ (mm) & 1.05 & 2.20 & 3.00 & 5.00 & 7.05 & 8.93 & 10.00 & 4.20 & 10.36 & 14.38\\
$\frac{\theta_a+\theta_r }{2}$ (deg) & 55 & 55 & 39 & 33 & 30 & 32 & 25 & 89 & 92 & 92 \\
$\frac{\theta_a-\theta_r }{2}$ (deg) & 13 & 15 &  9 & 15 & 5 & 15 & 10 & 11 & 13.4 & 13.4 
\end{tabular}
\end{ruledtabular}
\caption{\label{tab:tube} Diameters and air-water contact angles of our cylinders.}
\end{table*}

\section{Modeling}
\subsection{Mass and momentum conservation}
As the rivulet flows down, it is subjected to three forces: basal viscous friction, gravity and the inertial-capillary depression in the stream. The steady momentum balance on an immobile control volume (C.V) bounded by a control surface (C.S) is:  
\begin{align*}
\varoiint_\mathrm{C.S}\rho \boldsymbol{u} &\left(\boldsymbol{u}\cdot\boldsymbol{\hat{\mathbf{n}}}_\mathrm{C.S}\right) \mathrm{d}S= \\
&\iiint_\mathrm{C.V} \rho\boldsymbol{g} \mathrm{d}V  +\varoiint_\mathrm{C.S} \left(\boldsymbol{\tau}^* - P\, \boldsymbol{\hat{\mathbf{n}}}_\mathrm{C.S}\right) \mathrm{d}S,
\end{align*}
with $\rho$ the fluid density, $\boldsymbol{u}$ its velocity, $\boldsymbol{\tau}^*$ the shear stress, $P$ the pressure and $\boldsymbol{\hat{\mathbf{n}}}_\mathrm{C.S}$ the C.S normal unit vector. If we apply it to an infinitesimal section of the rivulet between $s$ and $s+\mathrm{d}s$ [as shown in Fig 2(a)], since the shear stress is non-zero only on the solid surface and $\boldsymbol{u}$ is aligned with the inlet and outlet normal (which is the rivulet centerline tangent unit vector $\boldsymbol{\hat{\mathbf{t}}}$), it simplifies to
\begin{align*}
&\rho\boldsymbol{\hat{\mathbf{t}}}(s+\mathrm{d}s) \iint_{A(s+\mathrm{d}s)} u^2 \mathrm{d}S - \rho\boldsymbol{\hat{\mathbf{t}}}(s) \iint_{A(s)} u^2 \mathrm{d}S=\\
& \mathrm{d}s\left[\rho \boldsymbol{g} A(s)  + \boldsymbol{\hat{\mathbf{t}}}(s) \int_{W(s)}\tau^* \mathrm{d}\ell - \oint_{\mathcal{C}} P\, \boldsymbol{\hat{\mathbf{n}}}_\mathrm{C.S}\mathrm{d}\ell\right].
\end{align*}
Here $A$ is the area of the rivulet cross section, $\mathcal{C}$ its perimeter and $W$ the length of the solid-liquid contact (which is the width of the rivulet for wetting solids). The perimeter of the pressure term $\mathcal{C}$ can be decomposed in the solid-liquid portion $W$ and the free surface (F.S). Since the rivulet is symmetric the resultant will be along the cylinder normal $\boldsymbol{\hat{\mathbf{n}}}$ and we rewrite it
\begin{align*}
\oint_{\mathcal{C}} P \boldsymbol{\hat{\mathbf{n}}}_\mathrm{C.S}\mathrm{d}\ell&=\boldsymbol{\hat{\mathbf{n}}}\left[\int_{W(s)}P\mathrm{d}\ell-\int_{\mathrm{F.S}}P\mathrm{d}\ell\right]\\
&= W\Delta P\, \boldsymbol{\hat{\mathbf{n}}}.
\end{align*}
Now if we assume a constant speed $u$ equals to its average over the cross section $U=\frac{1}{A}\iint_{A} u \mathrm{d}S$ and average the shear stress $\tau=\frac{1}{W(s)}\int_{W(s)} \tau^* \mathrm{d}\ell$ we get
\begin{equation*}
\rho\frac{\mathrm{d}\left(AU^2 \boldsymbol{\hat{\mathbf{t}}}\right)}{\mathrm{d}s}= W \left( \tau \, \boldsymbol{\hat{\mathbf{t}}} - \Delta P \, \boldsymbol{\hat{\mathbf{n}}} \right)+\rho A\mathbf{g}.
\end{equation*}
Applying mass conservation $Q=A(s)U(s)$ finally reduces the momentum balance to its form in the main text [Eq. (1)].

\subsection{Application to the helix}
In the helical rivulet regime, as the jet completely sticks to the cylinder the position vector of the trajectory can be written $\boldsymbol{r}(s)=\mathcal{R}_{\mathrm{h}}\cos\phi(s)\boldsymbol{\hat{\mathbf{e}}_x}+\mathcal{R}_{\mathrm{h}}\sin\phi(s)\boldsymbol{\hat{\mathbf{e}}_y}+z(s)\boldsymbol{\hat{\mathbf{e}}_z}$ with $\mathcal{R}_{\mathrm{h}}$ the (constant) helix radius, $\phi$ the azimuthal angle and $s$ the arc-length. The tangent vector is then 
\[\boldsymbol{\hat{\mathbf{t}}}(s)\equiv\frac{\mathrm{d}\boldsymbol{r}(s)}{\mathrm{d}s}=\mathcal{R}_{\mathrm{h}}\frac{\mathrm{d}\phi}{\mathrm{d}s}\left(-\sin\phi(s)\boldsymbol{\hat{\mathbf{e}}_x}+\cos\phi(s)\boldsymbol{\hat{\mathbf{e}}_y}\right)+\frac{\mathrm{d}z}{\mathrm{d}s}\boldsymbol{\hat{\mathbf{e}}_z}\]

The unwrapped 2D coordinate system ($X$,$z$) presented in Fig. 2 naturally appears by rewritting the tangent vector as $\boldsymbol{\hat{\mathbf{t}}}(s)=\frac{\mathrm{d}X}{\mathrm{d}s}\boldsymbol{\hat{\mathbf{e}}_X}+\frac{\mathrm{d}z}{\mathrm{d}s}\boldsymbol{\hat{\mathbf{e}}_z}$ with $X(s)=\mathcal{R}_{\mathrm{h}}\phi(s)$ and $\boldsymbol{\hat{\mathbf{e}}_X}=\boldsymbol{\hat{\mathbf{e}}_\phi}=-\sin\phi(s)\boldsymbol{\hat{\mathbf{e}}_x}+\cos\phi(s)\boldsymbol{\hat{\mathbf{e}}_y}$. The later can then be fully expressed with only the local angle $\psi(s)$ of the tangent vector with respect to the vertical as $\frac{\mathrm{d}z}{\mathrm{d}s}=\cos\psi(s)$ and $\frac{\mathrm{d}X}{\mathrm{d}s}=\sin\psi(s)$.

If we now look at the derivative of the speed in the full momentum balance Eq. (1), it gives in this geometry
\begin{align*}
\frac{\mathrm{d}\left(U\boldsymbol{\hat{\mathbf{t}}}\right)}{\mathrm{d}s}&=\frac{\mathrm{d}U}{\mathrm{d}s}\boldsymbol{\hat{\mathbf{t}}}(s)+U(s)\frac{\mathrm{d}\boldsymbol{\hat{\mathbf{t}}}(s)}{\mathrm{d}s}\\ 
&=\left(\frac{\mathrm{d}U}{\mathrm{d}s}\sin\psi(s)+U(s)\frac{\mathrm{d}\psi}{\mathrm{d}s}\cos\psi(s)\right)\boldsymbol{\hat{\mathbf{e}}_X}\\ 
&+\left(\frac{\mathrm{d}U}{\mathrm{d}s}\cos\psi(s)-U(s)\frac{\mathrm{d}\psi}{\mathrm{d}s}\sin\psi(s)\right)\boldsymbol{\hat{\mathbf{e}}_z}\\ 
&-U(s)\frac{\mathrm{d}\phi}{\mathrm{d}s}\sin\psi(s)\boldsymbol{\hat{\mathbf{e}}_r}.
\end{align*}

The last term is purely radial and thus in the normal direction as $\boldsymbol{\hat{\mathbf{e}}_r}=\boldsymbol{\hat{\mathbf{n}}}$ in this geometry. Therefore, it can be interpreted as a centrifugal force balancing the pressure term in Eq. (1). Balancing the two other terms in the $X$ and $z$ direction with the friction force and gravity using the assumption made in the main text [$\tau W=-3C\eta D_j^2 U^2/Q$] yields
\begin{align*}
\frac{\mathrm{d}U}{\mathrm{d}s}\sin\psi+U\frac{\mathrm{d}\psi}{\mathrm{d}s}\cos\psi&=-\frac{3\eta C D_j^2 U^2}{\rho Q^2}\sin\psi, \\
\frac{\mathrm{d}U}{\mathrm{d}s}\cos\psi-U\frac{\mathrm{d}\psi}{\mathrm{d}s}\sin\psi&=-\frac{3\eta C D_j^2 U^2}{\rho Q^2}\cos\psi+\frac{g}{U}. 
\end{align*}
The final equations [Eqs. (2)(3)] are then obtained by isolating the rivulet speed and angle and by replacing the flowrate with the initial jet speed $Q=U_0\pi D_j^2/4$.

\subsection{Application to the jet bending}
\begin{figure}[tb]
\centering
\includegraphics[width=0.8\columnwidth]{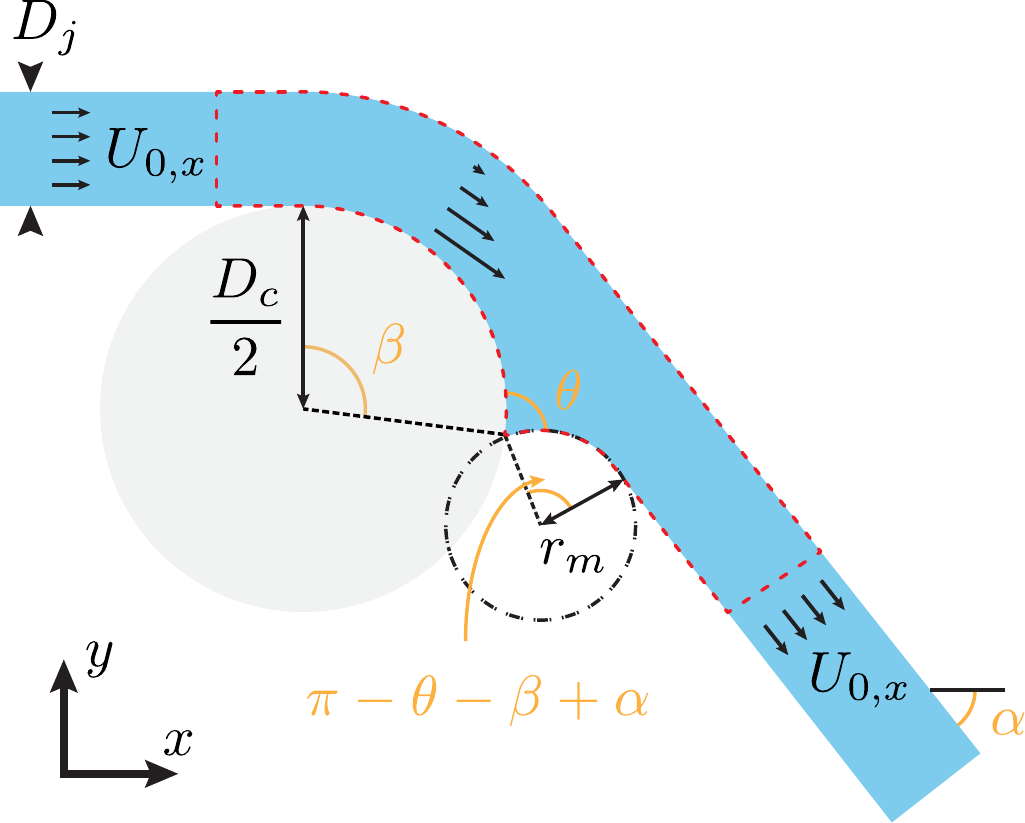}
\caption{\label{fig:schemforces} Schematic of model defining all quantities required for the calculation of the forces acting on the control volume (dashed red region). Angles are highlighted in orange.
}
\end{figure}

\paragraph{Two dimensional reduction.--}~In the jet bending regime, as the jet velocity is higher, we neglect gravitational and frictional forces and the flow is assumed steady, laminar, irrotational and inviscid. A more detailed sketch of the projected problem is shown in Fig. \ref{fig:schemforces}. We call $\alpha$ the angle (with respect to the horizontal $x$) at which the jet separate from the cylinder and $\beta$ the angle that the wetted portion of the cylinder makes. At the separation point a capillary meniscus, assumed circular with a radius $r_m$, is formed to accommodate the solid-liquid-air contact angle $\theta$ as shown by \citet{Duez:2010} (see Figs. 4(a) and \ref{fig:jetshape}). Only the pressure terms remains in the momentum balance:
\[\rho Q \frac{\mathrm{d}\left(U\boldsymbol{\hat{\mathbf{t}}}\right)}{\mathrm{d}s}=-\boldsymbol{\hat{\mathbf{n}}}W\Delta P.\]
Since only the pressure force (which is purely normal and thus have no $z$ component) remains the jet angle is constant $\psi(s)=\psi_0$ and $U(s)=U_0$. It is then convenient to integrate the momentum balance in the control volume whose projection is shown in Fig. \ref{fig:schemforces}:
\begin{align*}
&\int_{s_{\rm in}}^{s_{\rm out}} \rho Q U_0 \frac{\mathrm{d}\boldsymbol{\hat{\mathbf{t}}}}{\mathrm{d}s'} \mathrm{d}s'=\rho Q U_0 \left[\boldsymbol{\hat{\mathbf{t}}}(s_{\rm out})-\boldsymbol{\hat{\mathbf{t}}}(s_{\rm in})\right]\\
&=\rho Q U_0 \sin\psi_0 \left[\left(\cos\alpha-1\right)\boldsymbol{\hat{\mathbf{e}}_x}+\sin\alpha\boldsymbol{\hat{\mathbf{e}}_y}\right]\\
&=-\int \mathrm{d}s'\, \Delta P\, W\,\boldsymbol{\hat{\mathbf{n}}}
\end{align*}

If we now assume for simplicity a constant square cross section $A(s)=W(s)^2=D_j^2$, the pressure is uniform along the width and we can reduce the integral over the whole control surface (C.S) to a line integral over the top $\ell^{\rm top}$ and bottom $\ell^{\rm bot}$ section of the C.S.
\begin{align*}
\int \mathrm{d}s'\, \Delta P\,W\, \boldsymbol{\hat{\mathbf{n}}}=W\left[\int_{\ell^{\rm top}} \mathrm{d}s' P_{\rm top} \boldsymbol{\hat{\mathbf{n}}}+\int_{\ell^{\rm bot}} \mathrm{d}s' P_{\rm bot}\boldsymbol{\hat{\mathbf{n}}}\right].
\end{align*}
Since the inclination angle $\psi_0$ is constant, this line integral can then be projected line in the ($x$,$y$) plane as $\mathrm{d}s'=\mathrm{d}s_ {\parallel}'/\sin\psi_0$, with $\mathrm{d}s_ {\parallel}'=\sqrt{\mathrm{d}x'^2+\mathrm{d}y'^2}$. We finally obtain
\begin{equation*}
\rho U_{0,x}^2 D_j \left[\left(\cos\alpha-1\right)\boldsymbol{\hat{\mathbf{e}}_x}+\sin\alpha\boldsymbol{\hat{\mathbf{e}}_y}\right]=\int_{\ell_{\parallel}^{\rm top}+\ell_{\parallel}^{\rm bot}} P \boldsymbol{\hat{\mathbf{n}}} \mathrm{d}s_{\parallel}'
\end{equation*}
with $U_{0,x}=U_0 \sin\psi_0$ the projected initial speed in the $x$ direction. The full 3D momentum equation can therefore be projected in 2D and becomes equivalent to the problem presented in Fig. \ref{fig:schemforces}, which was also considered in \cite{Bouwhuis:2015}.

\paragraph{Pressure calculation.--}~The pressure differs from the atmospheric pressure in the curved region of the line integrals only (all pressure terms are expressed relative to the atmospheric pressure). In the curved part of $\ell_{\rm top}$ and in the meniscus region of $\ell_{\rm bot}$ there is a capillary pressure jump which creates forces of capillary origin $\boldsymbol{F}_{\rm top}$ and $\boldsymbol{F}_{\rm men}$. In addition the bending of the streamlines in the fluid create a depression which creates the hydrodynamic force $\boldsymbol{F}_{\rm hyd}$ on the portion of $\ell_{\rm bot}$ in contact with the cylinder. The momentum balance projected in both $x$ and $y$ directions can be written:
\begin{align}
\rho U_{0,x}^2 D_j \left(\cos\alpha-1\right) &= F_{\mathrm{hyd},x} + F_{\mathrm{top},x} + F_{\mathrm{men},x}, \label{eq:xmom} \\
-\rho U_{0,x}^2 D_j \sin\alpha  &= F_{\mathrm{hyd},y} + F_{\mathrm{top},y} + F_{\mathrm{men},y}.  \label{eq:ymom}
\end{align}
We now calculate these forces separately by integrating the fluid pressure over the relevant line integrals (boundaries of the control volume).

At the top interface in the curved circular region there is a first overpressure of capillary origin $P_{\rm top}=\gamma/\left(\frac{D_c}{2}+D_j\right)$. The top capillary force is then: 
\begin{align}
F_{\mathrm{top},x}&=-\int_0^\alpha  P_{\rm top} \left(\frac{D_c}{2}+D_j\right) \sin\beta^* \mathrm{d}\beta^* \nonumber \\
&=\gamma\left(\cos\alpha-1\right), \label{eq:topFx}\\
F_{\mathrm{top},y}&=-\int_0^\alpha P_{\rm top} \left(\frac{D_c}{2}+D_j\right) \cos\beta^* \mathrm{d}\beta^* \nonumber \\
&=-\gamma\sin\alpha. \label{eq:topFy}
\end{align}
Similarly within the meniscus region, $P_{\mathrm{men}}=-\gamma/r_m$ and knowing the meniscus angle $\pi-\beta-\theta+\alpha$ from geometry 
\begin{align}
F_{\mathrm{men},x}&=\int_{-\left(\pi-\beta-\theta\right)}^\alpha P_{\mathrm{men}} r_m \sin\beta^* \mathrm{d}\beta^* \nonumber \\
&=\gamma\left(\cos\left(\beta+\theta\right)+\cos\alpha\right), \label{eq:menFx}\\
F_{\mathrm{men},y}&=\int_{-\left(\pi-\beta-\theta\right)}^\alpha P_{\mathrm{men}} r_m \cos\beta^* \mathrm{d}\beta^* \nonumber \\
&=-\gamma\left(\sin\left(\beta+\theta\right)+\sin\alpha\right). \label{eq:menFy}
\end{align}

On the wetted solid surface there is an additional depression of hydrodynamic origin. To compute the pressure distribution in the liquid we must know the velocity profile $u(r)$ inside the flowing liquid (with $r$ the radial coordinate), for a given geometry characterized by $\tilde{R}$. As explained in the main text, we focus on the case $\tilde{R}\gg 1$ for which we expect from potential flow theory $u \sim 1/r$. The prefactor is then determined by mass conservation in the jet:
\[U_{0,x} D_j=\int_{D_c/2}^{D_c/2+D_j} u(r)\mathrm{d}r \quad \Rightarrow\quad u=\frac{U_{0,x}D_j}{r\ln\left(1+1/\tilde{R}\right)}. \]

The depression on the cylinder surface $P_{c}$ is thus:
\begin{align}
P_{c} &= P_{\rm top}-\rho\int_{D_c/2}^{D_c/2+D_j} \frac{u(r)^2}{r} \mathrm{d}r \nonumber\\
&=\frac{\gamma /D_j}{1+1/\tilde{R}} - \rho U_{0,x}^2 \frac{\mathcal{G}(\tilde{R})}{\tilde{R}} \nonumber
\end{align}
with $\mathcal{G}(\tilde{R})=\left(\frac{1}{2}+\tilde{R}\right) \left(\tilde{R}\left(1+\tilde{R}\right)^2\left[\ln(1+\frac{1}{\tilde{R}})\right]^2\right)^{-1}$ encoding the information about the velocity profile: assuming a different velocity profile simply modifies $\mathcal{G}(\tilde{R})$. We finally obtain the hydrodynamic force by integration of the pressure over the wetted area:
\begin{align}
F_{\mathrm{hyd},x}&=\int_0^\beta P_{c} \frac{D_c}{2} \sin\beta^* \mathrm{d}\beta^* \label{eq:hydFx}\\
&= \left[\rho U_{0,x}^2 D_j\mathcal{G}(\tilde{R})-\gamma\left(\frac{\tilde{R}}{1+\tilde{R}}\right)\right]\left(\cos\beta-1\right)\nonumber
\end{align}
\begin{align}
F_{\mathrm{hyd},y}&=\int_0^\beta P_c \frac{D_c}{2} \cos\beta^* \mathrm{d}\beta^* \label{eq:hydFy}\\
&=-\left[\rho U_{0,x}^2 D_j\mathcal{G}(\tilde{R})-\gamma\left(\frac{\tilde{R}}{1+\tilde{R}}\right)\right]\sin\beta. \nonumber
\end{align}
It is instructive to note that if there is no flow within the meniscus, by continuity $P_c=P_{\rm men}$ and the flow sets the meniscus size 
\[\frac{r_m}{D_j}=\frac{1+\tilde{R}}{\mathrm{We}_{\parallel}\left(1+\frac{1}{\tilde{R}}\right)\mathcal{G}(\tilde{R})-1}.\]

Finally, replacing the forces given by equations \eqref{eq:topFx} to \eqref{eq:hydFy} into the momentum balance \eqref{eq:xmom}\eqref{eq:ymom} and switching to dimensionless numbers finally leads to the final equations (see also \cite{Bouwhuis:2015}):

\begin{widetext}
\begin{align}
\left[\mathrm{We}_{\parallel}-\mathrm{We}_{\parallel}\mathcal{G}(\tilde{R})-\frac{1}{\left(1+\tilde{R}\right)}\right] + \left[2-\mathrm{We}_{\parallel}\right]\cos\alpha &+ \left[\mathrm{We}_{\parallel}\mathcal{G}(\tilde{R})-\frac{\tilde{R}}{\left(1+\tilde{R}\right)}\right]\cos\beta 
+ \cos(\beta+\theta) = 0, \label{eq:xcomp} \\
\left[2-\mathrm{We}_{\parallel}\right]\sin\alpha &+ \left[\mathrm{We}_{\parallel}\mathcal{G}(\tilde{R})-\frac{\tilde{R}}{\left(1+\tilde{R}\right)}\right]\sin\beta + \sin(\beta+\theta) = 0. \label{eq:ycomp}
\end{align}
\end{widetext}

\begin{figure}[tb]
\centering
\includegraphics[width=\columnwidth]{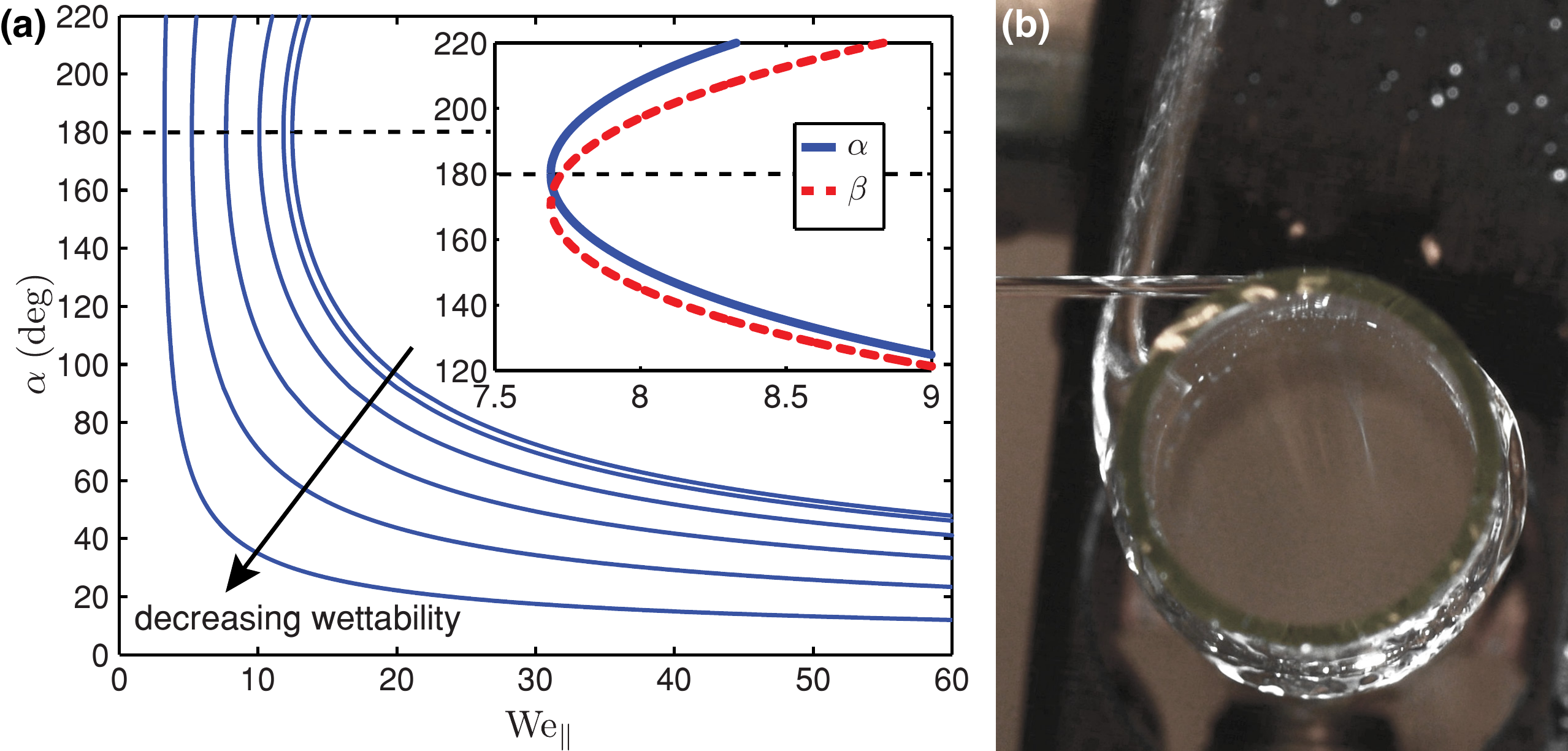}
\caption{\label{fig:Jet_dev} \textbf{(a)} Deviation angle $\alpha$ as a function of the projected Weber number $\mathrm{We}_{\parallel}$ for $\tilde{R}=4$ and various wettabilities $\theta=0,30,60,90,120,150^\circ$. \textit{Inset:} Detailed view of $\alpha$ and $\beta$ around the critical point for $\theta=90^\circ$. \textbf{(b)} Experimental view of the upper branch of solution with $\alpha>180^\circ$.}
\end{figure}

\paragraph{Analysis of the model.--}~%Four interesting asymptotic limits can be identified from Eqs. \eqref{eq:xcomp}\eqref{eq:ycomp}. For infinitely fast flows, $\mathrm{We}_{\parallel}\rightarrow\infty$, the only solution is $\alpha=\beta=0$, corresponding to no deviation (whatever the value of $\theta$). The same holds for perfectly non-wetting surfaces, i.e. $\theta\rightarrow 180^\circ$, for any $\mathrm{We}_{\parallel}$. On the contrary, for infinitely large cylinders, i.e. $\tilde{R}\rightarrow\infty$, no separated solution exists (unless $\theta=180^\circ$) and the jet sticks to the surface. A final special case is $\mathrm{We}_{\parallel}=2$ where the $\alpha$ dependence completely drops out of the equations, and we are left with two equations for a single unknown $\beta$. This has no solution unless $\theta=180^\circ$. As this indeed coincides with the Taylor-Culick velocity we can interpret $\mathrm{We}_{\parallel}=2$ as a minimum flow speed needed for a non-retracting sheet \cite{Taylor:1959,Culick:1960}. All these limiting cases are consistent with our experimental results.
We solve the momentum balance \eqref{eq:xcomp}\eqref{eq:ycomp} numerically and plot in Fig.~\ref{fig:Jet_dev}(a) the separation angle $\alpha$ as a function of $\mathrm{We}_{\parallel}$ for a typical cylinder radius $\tilde{R}=4$ at various wettabilities. Solutions only exist above a critical value of the Weber number, $\mathrm{We}_{\parallel c}$, which we identify as the threshold for the sticking transition. The critical point is found to coincide with $\alpha=180^\circ$. Above $\mathrm{We}_{\parallel c}$, the momentum balance admits two possible solutions. However, solutions for $\alpha$ larger than $180^\circ$ are only observed around the critical point [see Fig. \ref{fig:Jet_dev}(b)] and are experimentally unstable, thus suggesting a saddle node bifurcation with only the lower branch as stable. Now focusing on the lower branch, as expected, the deflection angle $\alpha$ increases when the jet speed is reduced, with much more rapid variations around the critical point following the saddle node scaling [$\alpha_c-\alpha\sim\left(\mathrm{We}_{\parallel}-\mathrm{We}_{\parallel c}\right)^{1/2}$]. The inset of Fig. \ref{fig:Jet_dev}(a) shows a zoom around the critical point for both angles, $\alpha$ and $\beta$. The two angles always take similar values, with a maximum difference of about $20^\circ$. The global minimum of $\beta$ is also reached at $\mathrm{We}_{\parallel c}$, but has a value slightly below $180^\circ$.

\begin{figure}[tb]
\centering
\includegraphics[width=\columnwidth]{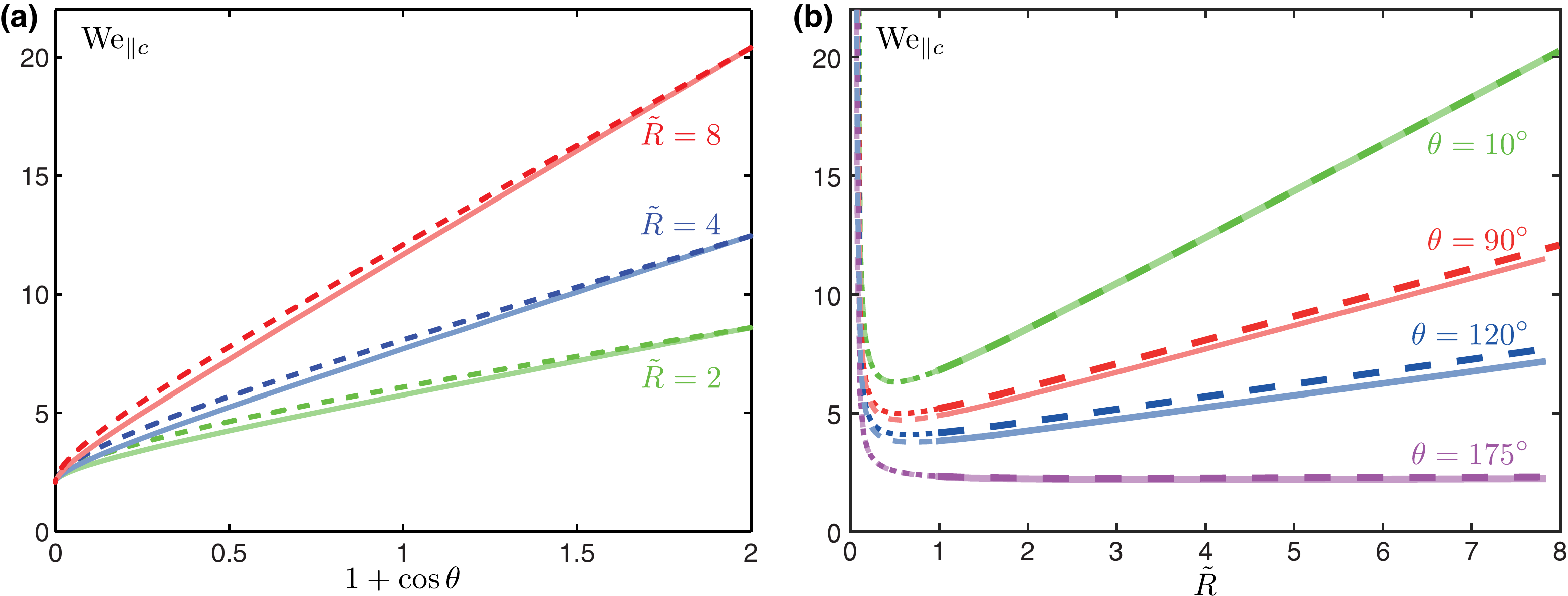}
\caption{\label{fig:arXiv_fig56} Critical projected Weber number $\mathrm{We}_{\parallel c}$ as a function of $1+\cos\theta$ for fixed values of $\tilde{R}$ \textbf{(a)} and as a function of $\tilde{R}$ for fixed values of $\theta$ \textbf{(b)}. Lighter solid curves: numerical solution. Darker dashed curves: analytical asymptotic expansion. The regime $\tilde{R}<1$ where the model assumption become invalid is highlighted with thinner dashes.}
\end{figure}

\begin{figure}[t]
\begin{minipage}{\linewidth}
\centering
\includegraphics[width=\columnwidth]{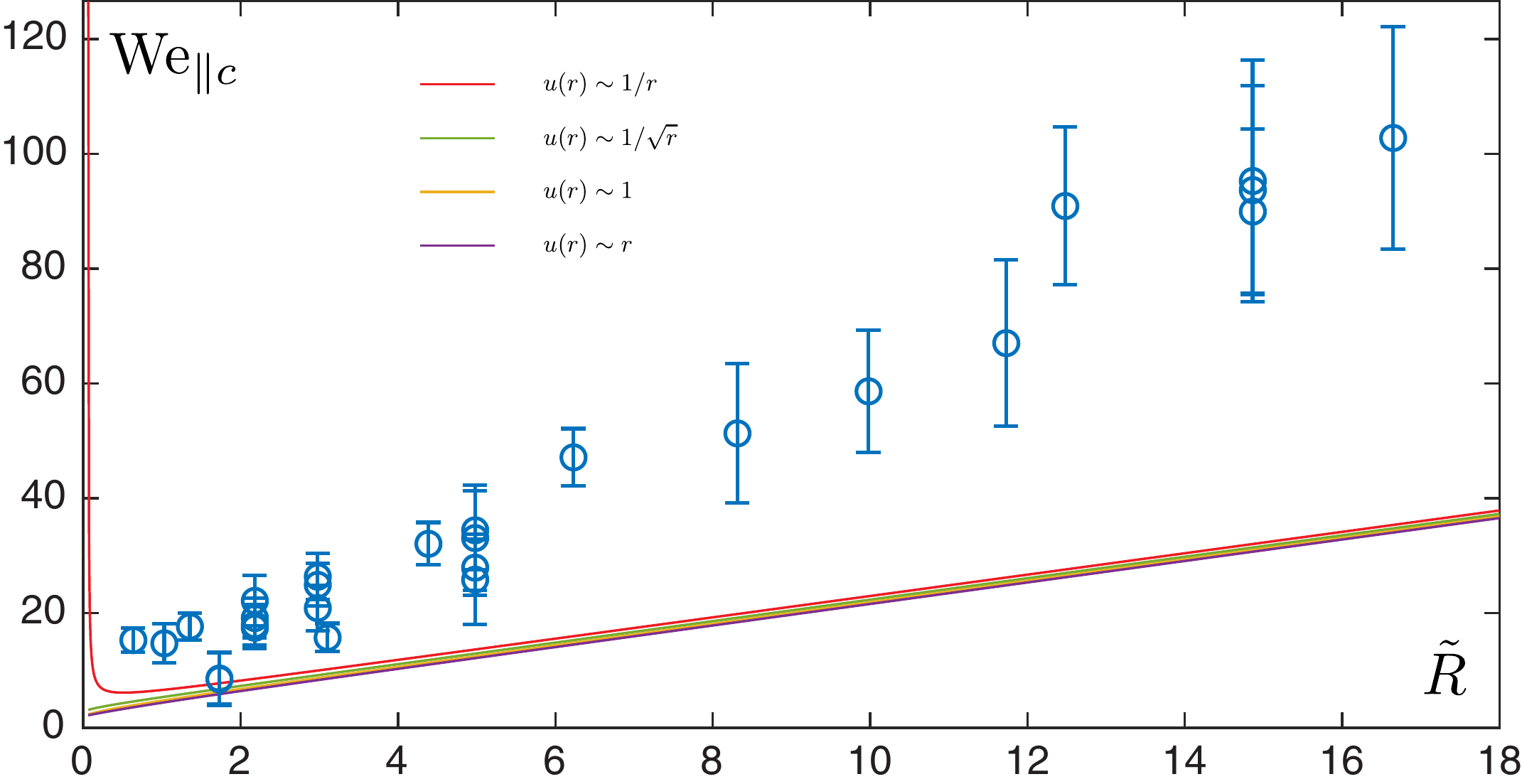}
\captionof{figure}{\label{fig:diffG} Influence of the shape of the velocity profile on the critical Weber number $\mathrm{We}_{\parallel c}$. Blue circles are the same data as Fig. 4(e), the solid curves are the full numerical solution of the momentum balance with $\theta=30^\circ$ for the different profiles shown in Table \ref{tab:Gfunc}.
}

\begin{ruledtabular}
\begin{tabular}{cc}
$u(r)$ & $\mathcal{G}(\tilde{R})$ \\
$\sim 1/r$ & $\left(\frac{1}{2}+\tilde{R}\right)/\left(\tilde{R}\left(1+\tilde{R}\right)^2\left(\ln(1+\frac{1}{\tilde{R}})\right)^2\right)$   \\
$\sim 1/\sqrt{r}$ & $1/\left(4(1+\tilde{R})(\sqrt{\tilde{R}}-\sqrt{1+\tilde{R}})^2\right)$\\
$\sim 1$ & $\tilde{R} \ln\left(1+\frac{1}{\tilde{R}}\right)$\\
$\sim r$ & $\tilde{R}/\left(1+\frac{\tilde{R}}{2}\right)$
\end{tabular}
\end{ruledtabular}
\captionof{table}{\label{tab:Gfunc} Velocity profile encoding function $\mathcal{G}(\tilde{R})$ for different power law velocity profiles  $u(r)\sim r^x$ (the prefactor is determined from mass conservation).}
\end{minipage} 
\end{figure}

\begin{figure}[tb]
\centering
\includegraphics[width=\columnwidth]{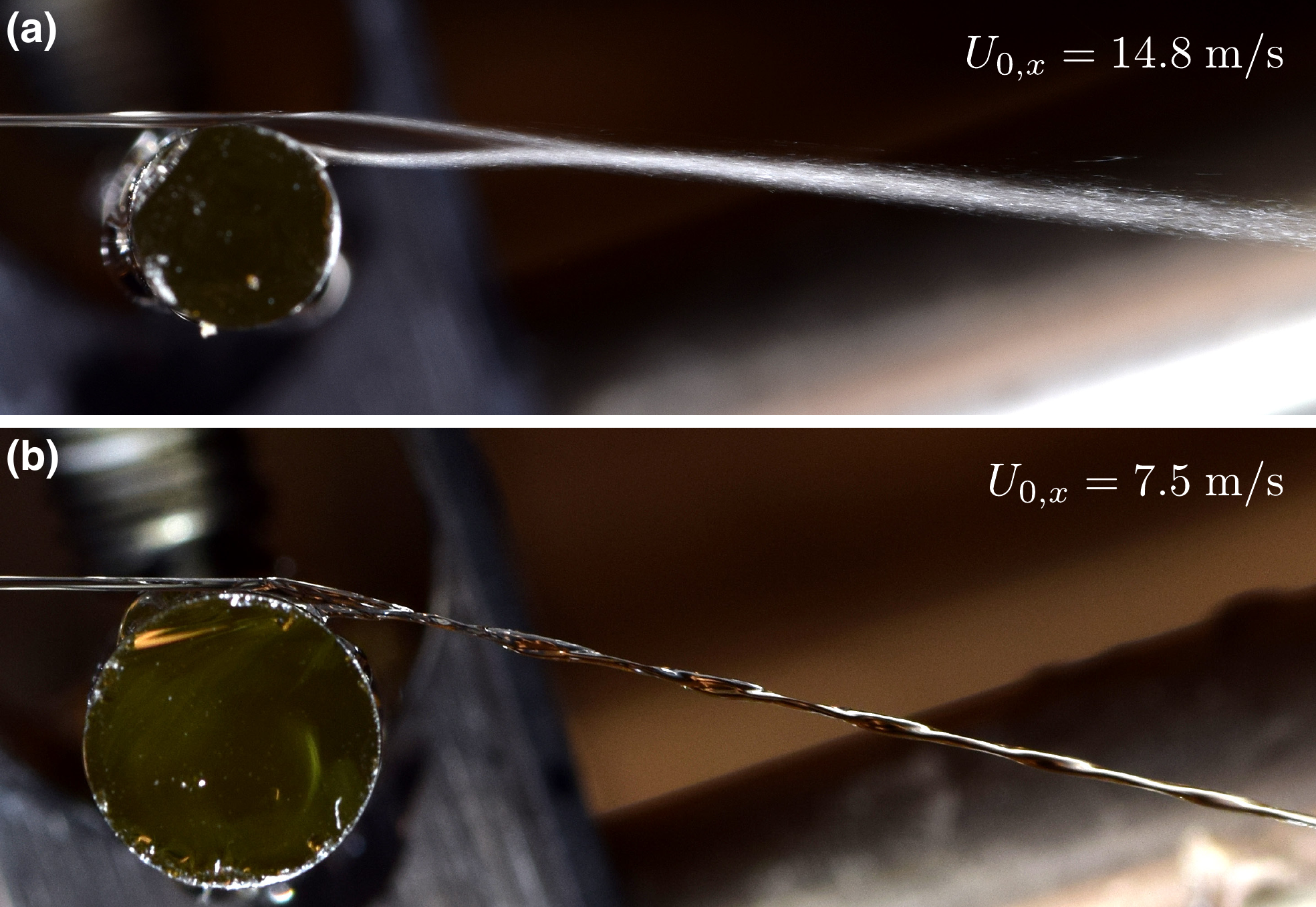}
\caption{\label{fig:jetshape} Morphology of the jet leaving the cylinder at high speed ($D_j=0.3$ mm, $\psi=90^\circ$). \textbf{(a)} At very high velocity the jet breaks into a stream of droplet (with well defined angle) after leaving the cylinder ($D_c=5.0$ mm). \textbf{(b)} At intermediate velocities the jet does not break but exhibits chain like undulations ($D_c=7.05$ mm).
}
\end{figure}

The critical Weber number depends on both the wettability of the solid and the curvature of the cylinder. Fig. \ref{fig:arXiv_fig56}(a) illustrates the wettability dependence: $\mathrm{We}_{\parallel c}$ varies linearly with  $1+\cos\theta$ as in \citet{Duez:2010} scaling analysis, except in the superhydrophobic limit ($\theta\approx 180^\circ$) where $\mathrm{We}_{\parallel c} \rightarrow 2$ for all cylinder radii. Fig. \ref{fig:arXiv_fig56}(b) illustrates the solid edge curvature dependence: for large $\tilde{R}$, the trend is also linear as in our experiments, a result which differs notably from the previous scaling analysis which found a quadratic dependence \cite{Duez:2010}. Note that the model is only valid for $\tilde{R}\geq 1$, owing to the assumptions of the velocity field. The model predicts an unphysical divergence for $\tilde{R}\ll 1$ that appears when $\mathcal{G}\rightarrow 1$. This divergence disappears when considering the more realistic corner solution $u(r)\sim r^{-\Omega}$ with $0<\Omega<1/2$ (see Fig. \ref{fig:diffG}).

\paragraph{Asymptotic expansion.--}~Since the sticking transition coincides with $\alpha=180^\circ$ and $\beta=180^\circ-\epsilon$ with $\epsilon$ a small parameter ($<20^\circ$), we can recover the critical Weber number through an asymptotic expansion around the critical point. Hence, we expand Eqs. \eqref{eq:xcomp}\eqref{eq:ycomp} around the critical point and linearize it up to the first order in $\epsilon$, i.e. replacing $\sin\epsilon\sim\epsilon$ and $\cos\epsilon\sim 1$. We find a quadratic equation for $\mathrm{We}_{\parallel c}$ such that:
\begin{align}
\mathrm{We}_{\parallel c} &= - A + \sqrt{A^2-B},\label{eq:asympsol}\\
\epsilon &= \frac{\sin\theta}{\mathcal{G}(\tilde{R})\mathrm{We}_{\parallel c}+\cos\theta-\frac{\tilde{R}}{1+\tilde{R}}}, 
\end{align}
with 
\begin{align}
A&=\frac{\frac{2\tilde{R}}{1+\tilde{R}}+(3\mathcal{G}(\tilde{R})-2)\cos\theta+\mathcal{G}(\tilde{R})\frac{3-\tilde{R}}{1+\tilde{R}}}{4\mathcal{G}(\tilde{R})(\mathcal{G}(\tilde{R})-1)}\\
B&=\frac{\frac{1-\tilde{R}}{(1+\tilde{R})^2}+\frac{3}{1+\tilde{R}}\cos\theta}{2\mathcal{G}(\tilde{R})(\mathcal{G}(\tilde{R})-1)}.
\end{align}

The analytical solution \eqref{eq:asympsol} is superimposed as dashed lines in Fig. \ref{fig:arXiv_fig56} and gives a very good description of the full numerical solutions. The difference is largest for $\theta\sim 90^\circ$, where the largest values for $\epsilon$ are encountered. We can even further simplify \eqref{eq:asympsol} when $\tilde{R}\gg 1$ as $\mathcal{G}(\tilde{R})=1-\frac{1}{2\tilde{R}}+o\left(\frac{1}{\tilde{R}^2}\right)$ which yields
\begin{equation}
\mathrm{We}_{\parallel c} \approx 4+ \tilde{R}\left(1+\cos\theta\right),
\label{eq:Wecexp}
\end{equation}
predicting a linear behavior of the critical Weber number with respect to both the aspect ratio $\tilde{R}$ and to $1+\cos\theta$.

\paragraph{Discussion of the assumptions.--}~
The first assumption is about the flow profile, the $u\sim 1/r$ profile is only valid for inviscid, irrotational flows around large cylinders. In reality, there must be some viscous effects with a boundary layer around the surface and we expect a slightly different flow structure \cite{Lhuissier:2012}. Besides, for small cylinders even in the inviscid case we expect a different solution: the corner flow. We thus calculated the hydrodynamic force for different power law profiles using the same procedure, which yielded different $\mathcal{G}(\tilde{R})$ (see Table \ref{tab:Gfunc}).
We replot in Fig. \ref{fig:diffG} the experimental $\mathrm{We}_{\parallel c}$ as a function of $\tilde{R}$ for our glass cylinders and compare it to the result of the model with the different velocity profiles of Table \ref{tab:Gfunc} (still with $\theta=30^\circ$). The choice of velocity profile has a minor influence on the critical Weber number except for $\tilde{R}\ll 1$ where the unphysical behavior $\mathrm{We}_{\parallel c}\rightarrow \infty$ as $\tilde{R}\rightarrow 0$ only appears for $u\sim 1/r$ and cannot explain the discrepancy with experiments. 
%It is also interesting to see that the previous scaling analysis for the pressure \cite{Duez:2010} corresponds to the linear velocity profile. 

%\begin{table*}
%\centering
%\begin{ruledtabular}
%\begin{tabular}{c|cccc}
%$u(r)$ & $\sim 1/r$ & $\sim 1/\sqrt{r}$ & $\sim 1$ & $\sim r$  \\
%$\mathcal{G}(\tilde{R})$  &  $\frac{\frac{1}{2}+\tilde{R}}{\tilde{R}\left(1+\tilde{R}\right)^2\left(\ln(1+\frac{1}{\tilde{R}})\right)^2}$ & $\left(4(1+\tilde{R})(\sqrt{\tilde{R}}-\sqrt{1+\tilde{R}})^2\right)^{-1}$ & $\tilde{R} \ln\left(1+\frac{1}{\tilde{R}}\right)$ & $\frac{\tilde{R}}{1+\frac{\tilde{R}}{2}}$
%\end{tabular}
%\end{ruledtabular}
%\caption{\label{tab:Gfunc} Velocity profile encoding function $\mathcal{G}(\tilde{R})$ for different power law velocity profiles  $u(r)\sim r^x$ (the prefactor is determined from mass conservation).}
%\end{table*}
%\begin{figure}[tb]
%\centering
%\includegraphics[width=\columnwidth]{Wec_diffG.pdf}
%\caption{\label{fig:diffG} Influence of the shape of the velocity profile on the critical Weber number $\mathrm{We}_{\parallel c}$. Blue circles are the same data as Fig. 4(e), the solid curves are the full numerical solution of the momentum balance with $\theta=30^\circ$ for the different profiles shown in Table \ref{tab:Gfunc}.
%}
%\end{figure}

The remaining assumptions are mostly geometric. 
Although in the model the jet is assumed coherent and square with a constant cross-section this is not the case in experiments. The jet is initially circular, then it makes a small hydraulic jump upon impact \cite{Kibar:2017} (it is widens with two rims and a thinner part in the center) that recoils and turn circular again after the separation point. Moreover, at very high speed, the jet leaving the cylinder breaks up in a stream of droplets with a well defined angle $\alpha$ [the one reported, see Fig. \ref{fig:jetshape}(a)] while at intermediates velocities although the jet does not break it forms a chain like structure similar to the one observed in pouring flows or colliding jets \cite{Bush:2004,Celestini:2010} [see Fig. \ref{fig:jetshape}(b) and the leftmost picture of Fig. 1(b)]. Finally, the meniscus is also often more complex than the simple circular picture (see Fig. \ref{fig:jetshape}). We thus believe that the discrepancies between the model and the experiments come from our oversimplified geometry. Though we expect the exact jet geometry to be only accessible through computational fluid dynamic simulations \cite{Kibar:2017} and therefore these assumptions to be hard to improve.

%\bibliography{bibspijet}
%merlin.mbs apsrev4-1.bst 2010-07-25 4.21a (PWD, AO, DPC) hacked
%Control: key (0)
%Control: author (8) initials jnrlst
%Control: editor formatted (1) identically to author
%Control: production of article title (-1) disabled
%Control: page (0) single
%Control: year (1) truncated
%Control: production of eprint (0) enabled
%